# USING HIGH PERFORMANCE COMPUTING AND MONTE CARLO SIMULATION FOR PRICING AMERICAN OPTIONS

Verche Cvetanoska
European University
Skopje, Macedonia

Toni Stojanovski
European University
Skopje, Macedonia

ABSTRACT

High performance computing (HPC) is a very attractive and relatively new area of research, which gives promising results in many applications. In this paper HPC is used for pricing of American options. Although the American options are very significant in computational finance; their valuation is very challenging, especially when the Monte Carlo simulation techniques are used. For getting the most accurate price for these types of options we use Quasi Monte Carlo simulation, which gives the best convergence. Furthermore, this algorithm is implemented on both GPU and CPU. Additionally, the CUDA architecture is used for harnessing the power and the capability of the GPU for executing the algorithm in parallel which is later compared with the serial implementation on the CPU. In conclusion this paper gives the reasons and the advantages of applying HPC in computational finance.

## I. INTRODUCTION

To calculate the value of American options numerically intensive methods must be applied: differential methods, fast Fourier transform and Monte Carlo simulations.

The Monte Carlo method is a numerical computational method commonly used in simulating physical problems, where it is impossible or impractical to obtain analytical (or closed form) solution for the system of equations. Using Monte Carlo is very convenient because the computational time of Monte-Carlo simulation increases approximately linearly with the number of variables, while in most other methods, the computational time increases exponentially with the number of variables.

American options represent a challenging problem in computational finance due to their early exercise feature. These options can be exercised at any time up to maturity.

In recent researches good results are achieved when American options are priced with Monte Carlo simulations. [2][6][7][11]

This paper is focused on using the capabilities of the graphics processing unit (GPU), and also will compare the performances between GPU and CPU when both are used for processing the same algorithm. Both GPU and CPU will be used for valuing American options using Quasi Monte Carlo simulations. Valuation will be made on a single option on a large number of stock pricing paths needed for higher accuracy of the Monte Carlo simulation.

## II. PROJECT GOALS

The purpose of this paper is to give a brief description of American options, and also to suggest a parallel way of solving them by using Quasi Monte Carlo simulations.
In this paper we will answer the following questions:
1. How to evaluate the American options in discrete time periods?
2. Which simulation to apply in order to achieve more accurate evaluation of the American options?
3. How to optimally use the limited resources of the GPU for valuing options?
4. How to use the SIMT method which is part of the CUDA architecture in order to evaluate the American options?
5. What is the speed up of the parallel implementation of the algorithm over the serial implementation?
6. What is the advantage of using GPGPU when the Monte Carlo simulation is used for American option pricing?

## III. BACKGROUND

### A. Why GPU computing?

With the development of multi-core processors the inability to process large amounts of data for a very short time was solved. Hardware architecture of graphic cards was the most convenient for this problem to be overcome. The graphics card has many cores and each core has hardware support for many threads. This construction of the graphics card was used for it to be applied as a graphics card for general purpose - GPGPU. [14]

NVidia, a well-known manufacturer for graphics cards, accommodated the graphics cards that appeared after the 8400 GS to be capable of processing a lot of data in parallel. In 2006 NVidia developed the CUDA architecture. The CUDA architecture uses the C/C++ language.

If the programmer knows how to properly use the CUDA architecture, and with proper allocation of the processes, very high system performance can be achieved [15]. Recent researches indicate that the GPU can process data up to ten times faster than the CPU [16]. When optimization on the CPU is not performed, then the GPU performances can be a hundred times better than the CPU performances [17].

### B. Options and Option pricing

Options are financial derivative instruments and represent a contract where the holder has the right but not the obligation to buy (or sell) an underlying asset for a determined price at the determined date. Options that give its holder the right to buy the underlying asset are called call options. Options when the holder gains the right to sell the underlying asset for a determined price at the determined date are called put options. Options exist in two main categories: vanilla options and exotic options.

In 1973 Fisher Black and Myron Scholes developed a closed form solution to price plain vanilla European Call/Put options. This closed form solution today is known as the Black-Scholes formula [1].

Depending on when the option can be exercised, there are different types of options, such as European, Bermudan and American Options.

*C. Black-Scholes model*

Black-Scholes model first appeared in 1972 as a title in the journal "Journal of Finance" in which Fischer Black and Myron Scholes did an empirical study. [2]

The Black-Scholes model is a model for pricing European options that can be exercised at the expiry date *T*. This model is the foundation for more complex models. This model provides a partial differential equation (PDE) for evaluation of the option price and is based on several assumptions, such as: the underlying asset price (spot price) follows a log normal distribution, the volatility of the underlying spot price is constant and the risk free rate of return is constant. [1]

*1) The Black Scholes pricing formulas*

The Black-Scholes formulas for pricing European call and put options at time *T*=0 are:

$$c = S_0 * CND(d_1) - X * e^{-r*T} * CND(d_2) \quad (1)$$

$$p = X * e^{-r*T} * CND(d_2) - S_{t_0} * CND(d_1). \quad (2)$$

$$d_1 = \frac{\log(\frac{S_0}{X}) + (r + \frac{v^2}{2})T}{v\sqrt{T}}. \quad (3)$$

$$d_2 = \frac{\log(\frac{S_0}{X}) + (r - \frac{v^2}{2})T}{v\sqrt{T}} = d_1 - v\sqrt{T}. \quad (4)$$

$$CND(-d) = 1 - CND(d). \quad (5)$$

where *c* is the value for a call option, *p* is the value of a put option, CND(*d*) is the Cumulative Normal Distribution function, $S_0$ is the current price of the underlying asset, *X* is exercise price, *T* is expiry time, *r* is continuously compounded risk free interest rate, *v* is implied volatility for the underlying asset.

*D. Monte Carlo and Quasi Monte Carlo*

Monte Carlo simulation is a popular technique for options pricing. Monte Carlo simulation is used for solving complex problems, such as high-dimensional integrals. While Monte Carlo is very useful for solving these major problems, it has one drawback, namely the need for great computing power. [3]

Monte Carlo simulation was invented by Stanislaw Ulam, Fermi Von Neumann and Metropolis in the 1940's. [4]. Monte Carlo simulation can be used for solving not just one but multiple problems. [18] An example of solving multiple problems at once is the pricing of options where at once thousands of options need to be priced. The Monte Carlo approach simulates paths for asset prices. Because many independent paths need to be calculated at once, it becomes clear that one of the most important features of the Monte Carlo simulation is the parallelization. The Monte Carlo simulation converges faster for more dimensional problems, requires less memory and its programming is easier than others techniques for pricing options.

The difference between Monte Carlo and Quasi Monte Carlo is that points in Monte Carlo ($x_1, x_2, x_3, ..., x_N$) are randomly chosen and independent, unlike the Quasi Monte Carlo simulation where points are generated quasi randomly. In many applications, this method proved to be a method with more advantages compared to the traditional Monte Carlo simulation method. This is due to its faster convergence and higher accuracy [5][6][7] [8] [9].

We use the Quasi Monte Carlo approach because it improves the convergence properties of the Monte Carlo techniques.

In our algorithm for pricing American options our purpose is to generate multiple samples for each path. We perform permutation on the quasi-random arrays by using linear congruential generator in order to generate statistically independent samples for each path. These samples are generated with uniform distribution by this quasi-random generator, and then they are normally distributed *N*(0, 1) by using the Moro Inverse Cumulative Normal Distribution.

When the Monte Carlo simulation is used for option pricing it actually estimates the expected *put* and *call* values. The core of this method is corresponding to the underlying Wiener process described in [1], generating *N* numeric samples using normal or pseudo normal distribution *N*(0,1). This process then averages the possible end period stock profits for every single path:

$$c_{mean}(S_t, T) = \frac{1}{N}\sum_{i=1}^{N} c_i(S_t, T) \quad (6)$$

$$p_{mean}(S_t, T) = \frac{1}{N}\sum_{i=1}^{N} p_i(S_t, T) \quad (7)$$

By discounting the estimated future price with $e^{-rT}$ we get an estimation of the present fair value of the derivative:

$$(p/c)_{fair}(S_t, 0) = (p/c)_{mean}(S_t, T)e^{-r*T} \quad (8)$$

Here *c* (call) and *p* (put) values are calculated by (6) and (7).

*E. American options*

The early methods used for pricing American options were binomial trees and other lattice methods, such as trinomial trees and finite difference methods. These methods were used to solve the partial differential equation (PDE) and its associated boundary value.

Unlike European options where the holder exercises the options at maturity, the American option gives the holder the right to choose when to exercise the options, that is, the holder has the right to exercise the option at any moment up to the maturity of the option.

American options always give higher value for the option than European options, because at least the final option value is equal to the European option price.

Pricing the American option with Monte Carlo simulation is very difficult because of the inability to approximate in continuous time and because of the American options early exercise feature. [9][10][11][12]

In order to avoid this problem, American options are considered as Bermudian options. In Bermudian options the whole time period is divided into several discrete times. For example, if American option can be exercised in any point of time $t$ where $0 \le t \le T$, in Bermudian style options, the option can be exercised only at a fixed set of times $0 \le t_1 \le t_2 \le t_3 ... \le t_n=T$.

In the American option pricing, the option holder compares the early exercise value and expected continuation value. This helps him to decide to early exercise the option or to keep it for the later time, when its value will be higher because his intention is to maximize the income, that is, to get the most optimal value for the given option.

Exercise boundary is an essential term in American option pricing. The exercise boundary refers to a threshold value for each time step, which shows if the option should be exercised or not.

For a call option, the holder should exercise the option when the value of the option is above the exercise boundary whereas for the put it is the opposite. The boundary at maturity is the exercise price, or the European option price.

Different methods make approximation of the exercise boundary in different ways. Simulation techniques like Monte Carlo try to find the optimal option value for every path where the value is above the exercise boundary, and then make the approximation.

Methods and techniques for pricing American options are described in section IV.

## IV. RELATED WORK

In [10], Broadie and Glasserman developed algorithm for pricing American- style securities using Monte Carlo with two branch processes. The first process gives an upper bound on the option price and the second gives a lower bound on the option price. The two processes converge to the true price.

Rogers in [11] also developed an algorithm for pricing American options using Monte Carlo simulation. He uses direct simulation approach, based on dual formulation on the optimal exercise problem. This method leads to an upper bound on the option price. Van Roy and Tsitsiklis [19] have introduced simulation-based methods for pricing the complex American option by iteration.

## V. ALGORITHM

The algorithm that we implemented is based on backward induction or dynamic programing principle. This approach is similar to [9] and [19], but the difference is that they don't calculate the option in the period [$t_{m-1}, t_m$] with the Black-Scholes formula

Our purpose by using this approach is to find the optimal exercise boundary by which we will get upper bound on the value of the American option.

Because of inability to estimate the American option price in continuous time, let's suppose that the American option is the Bermudian type of option with the possibility to exercise the option in discrete time steps $T = \{t_1, t_2, t_3, t_4, ..., t_m = T\}$.

Since there is no optimal strategy for exercising the option, we start from the period T and continue backward through all time points.

The price of the American option is at least equal to the European one. If the American option is exercised at expiry time $t_m=T$, then the price of the American option is equal to the price of the European option.

Expiry period is divided by $m$ equally spaced points. The distance between the points is $t=T/(m+1)$. The points are $t_i$ for $i=\{1,2,…,m\}$.

At the point $t_{i-1}$ the underlying stock is approximated by the Monte Carlo simulation

$$S_{t_{i-1}} = S_{(t_{i-1})-1} * e^{((r-\frac{1}{2}v^2)*t+v*\sqrt{t})(z_{i-1})} \quad (9)$$

We need the underlying asset value at time $t_{i-1}$ in order to calculate the exercise boundary in the period [$t_{i-1}, t_i$].

Here we make an assumption that at time $t_0$ the underlying price is the primary underlying price, and at time $t_i$ the value of the underlying asset is equal to the exercise value.

The calculation of the exercise boundary is performed by using the Black-Scholes formula $c(S_{t_{i-1}}, X, t, v, r)$ (1).

After calculating the value of the call option for the period [$t_{i-1}, t_i$] this value will be the exercise boundary for the previous point $t_{i-1}$. At time point $t_{i-1}$ the call value is calculated by:

$$c_{t_{i-1}} = \max(S_{t_{i-1}} - X, c(S_{t_{i-1}}, X, t, v, r)) \quad (10)$$

Continuing backward, the value of the call option is calculated by:

$$c_{t_{(i-1)-1}} = \max(S_{t_{(i-1)-1}} - X, c_{t_{i-1}} * e^{-r*(t)}) \quad (11)$$

Here $X$ is the exercise price, $r$ is the risk free rate of return, $v$ is the volatility, $t$ is the time point, and $z_i$ is the random generated sample for every time step. Then we approximate the final call value by using (6).

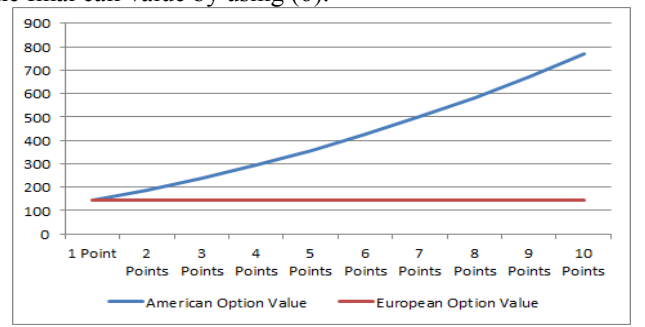

Figure 1: Value of American option vs. number of exercise points.

In this algorithm we assume that all future values of the underlying stock price are known, and we can choose the best time point to exercise the option. Therefore, the algorithm gives the upper bound the value of the American option.

Figure 1 shows the value of the American option as a function of the number of exercise points.

## VI. CUDA ARCHITECTURE AND PARALLEL IMPLEMENTATION

Developing a parallel algorithm is not an easy task to do. Our algorithm is based on the NVidia SDK samples for pricing the European Options with Monte Carlo simulation. In order to implement parallel and accurate algorithm for pricing American options with Quasi Monte Carlo simulation we modified and adopted these samples [16]. We must consider the CUDA resource constrains, such as memory bandwidth, number of blocks, threads etc. in order to implement an effective and fast algorithm.

First of all there are multiple types of CUDA memories, such as, registers, shared memory, and constant memory. These memories can be accessed at higher speed and in more parallel manner than the global memory.

The techniques which are very important for achieving the performance upgrading are titled as: Reduction technique, Global memory bandwidth, Dynamic partitioning of the SM resources, Data prefetching and Instruction mix

The reduction technique is very important when we want to achieve fast program execution and proper thread organization.

By using this technique divergence is avoid within the warp. That is accomplished by adding every element within shared memory with the element which is half section away from it. The size of the section is equal to the block size. All threads within the warp follow the same path for execution.

Divergence is avoided by using this way of calculating partial sums, but there is still a one problem, half of the elements remain idle after the first loop, and more and more after every loop. Solution to this problem is to halve the number of blocks and to replace them into the single load.

This means that we put the result from the one thread from global memory and the other thread from global memory which is half a section away from the first one into the shared memory directly. The second element is further half a section than the first element. This helps all threads within the blocks to be used.

Another technique very important is the instruction mix. Using this technique we can achieve performance improvements when we avoid loops, calculation using floating point number etc. Because the technique mentioned above uses loops for each iteration, in order to calculate the section size, the following change is performed. Because the maximum thread block size is 512, we calculated every possible block size and put that in the runtime.

```
if (blockSize >= 512) {
if (tid < 256) { add[tid] += add[tid + 256]; } __syncthreads();}
```

Also when the number of threads within the warp is smaller than 32, instructions within the warp are SIMD synchronous. So it is recommended to use the unrolling of the loop by defining all the possibilities for adding threads. When the number of threads is smaller than 32 it is recommended to avoid the method __syncthreads() because it can affect the performance.

The global memory bandwidth also had to be taken into account. In order to get maximum speed accesses to global memory, and also fast global memory response, we used the technique called coalescing. With this technique we load threads going through the column instead of loading from the row.

We also used the technique called dynamic partitioning of the SM resources. In our algorithm we check for the number of paths used. If the number is large, than smaller number of blocks will be used but larger number of threads. Because the number of registers required is large, the accommodated number of blocks will be small. By using this technique we make dynamic partitioning of the SM resources. This helps performance improvement to be achieved because the resources in SM are limited.

By using the samples from [20] and [16] we succeeded to implement an algorithm which works very fast and accurately.

The test results are made on a system which uses Intel Core i7 2.20 GHz processor, and on a graphics card NVidia GeForce GT 540 M. The GPU has 96 CUDA cores and is supported for using the CUDA architecture.

Our test results show that when more threads are used for observing huge number of paths then we get a significant increase in performance. If the number of blocks is higher than is necessary for performing the valuation, by the results below in the Table 2, it is easy to notice that it can cause degradation of the performance. It is noticeable that we need to know how to use the resource constrains in order to achieve a performance increase.

In the Table 2, below are listed the achieved increases in performance, which depend on the number of paths, number of threads per block, and the number of blocks per option, used for the Monte Carlo simulation.

In Table 2 below when the number of paths is smaller than 8192, then one block per option is used.

Table 2: GPU and CPU speed comparison.

| Threads Blocks Per option Speed | 256 64 CPU | 512 6 CPU | 256 64 GPU | 512 16 GPU | 512 64 GPU |
|---|---|---|---|---|---|
| 10 | 0.08 | 0.09 | 0.04 | 0.06 | 0.04 |
| 100 | 0.18 | 0.24 | 0.06 | 0.07 | 0.04 |
| 1000 | 1.27 | 1.23 | 0.09 | 0.04 | 0.06 |
| 10000 | 11.9 | 11.43 | 0.09 | 0.07 | 0.04 |
| 100000 | 110.62 | 110.24 | 0.13 | 0.08 | 0.16 |
| 200000 | 229.02 | 222.02 | 0.11 | 0.09 | 0.16 |
| 300000 | 337.91 | 335.34 | 0.12 | 0.11 | 0.20 |
| 500000 | 555.97 | 563.49 | 0.18 | 0.15 | 0.20 |
| 1000000 | 1100.15 | 1125.76 | 0.23 | 0.16 | 0.21 |

The results from Table 2 demonstrate that when the number of paths used for the simulation is small then the execution times on CPU and GPU are very close. For ten paths the CPU computational time is double the time than the GPU computational time. For hundreds of paths to thousands of paths the GPU is approximately three times faster than CPU in executing the algorithm. For 1000 to 10000 of paths the execution of the algorithm on the GPU is thirty time shorter than that of the algorithm executed on the CPU. For 10000 to 100000 GPU is more than 150 times faster than CPU when executing the algorithm. The maximum speedup is achieved when one million paths are used. In this case the CPU is more than 6500 times slower than GPU. As more paths are used the speed of CPU decreases, while the GPU speed remains very close as the time when only 10 paths were used.

Figure 2, shows the difference in the execution time of the same algorithm for American options in ten points on the CPU and the GPU.

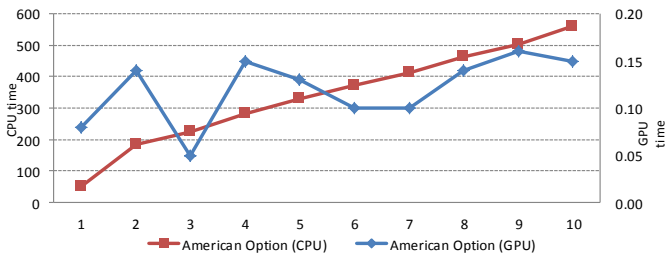

Figure 2: Difference in execution time on CPU and GPU.

From these results it is obvious that the GPU is the most suitable solution for today's needs of processing a huge amount of data for a very short time. As the number of simulation pats increases and the accuracy of results improves, execution time of Monte Carlo simulation doesn't degrade when executed on GPU.

## VII. CONCLUSION

Pricing of American Options using Monte Carlo simulation is an active area of academic research. In this paper we describe serial and parallel implementation of an algorithm for pricing of American options. The algorithm gives the upper bound on the value for the American options. The returned call value for equivalent input parameters on both CPU and GPU is the same and always higher than the European option with the same input parameters. We also implemented a very fast parallel algorithm, which is capable of simulating more than one million paths for Monte Carlo simulation in approximately 0.15µs.

We showed that the parallel algorithm implemented on the GPU is more than 6500 times faster than the serial one implemented on the CPU.

## REFERENCES

[1] Fischer Black and Myron Scholes, "The Pricing of Options and Corporate Liabilities," *Journal of Political Economy*, vol. 81, no. 3, pp. 637-654, May/June 1973.

[2] Don M. Chance. *Essays in Derivatives: Risk-Transfer Tools and Topics Made Easy*. John Wiley & Sons Inc., Hoboken, New Jersey, pp. 1-37, 2008.

[3] Quiyi Jia. "Pricing American Options using Monte Carlo Methods," Department of Mathematics, Uppsala University, June 2009.

[4] Wiliam A. Beyer, Peter H. Sellers and Michael S. Waterman. "Stanislaw M. Ulam's Contributions to Theoretical Theory," *Letters in Mathenlatical Physics, vol.*10, pp. 231-242, *1985*.

[5] Xiang Tian and Khaled Benkrid. "High Performance Quasi-Monte Carlo Financial Simulation: FPGA vs. GPP vs. GPU*," Journal ACM Transactions on Reconfigurable Technology and Systems (TRETS),* vol. 3, Issue 4, November 2010.

[6] Justin W.L. Wan, Kevin Lai, Adam W. Kolkiewicz and Ken Seng Tan. "A Parallel Quasi-Monte Carlo Approach to Pricing American Options on Multiple Assets*," International Journal of High Performance Computing and Networking,* vol. 4, No.5/6,  pp. 321 – 330, 2006.

[7] G. Larcher and G. Leobacher. "Quasi-Monte Carlo and Monte Carlo Methods and their Application in Finance," Department of Financial Mathematics, University Linz.

[8] Xiang Tian and Khaled Benkrid. "Massively Parallelized Quasi-Monte Carlo Financial Simulation on a FPGA Supercomputer," *Second International Workshop High-Performance Reconfigurable Computing Technology and Applications*, pp. 1 – 8, 2008.

[9] Eric Couffignals. "Quasi-Monte Carlo Simulations for Longstaff Schwartz Pricing of American Options*,"* Mathematical Institute, Lady Margaret Hall, University of Oxford, November 2010.

[10] Broadie M. and P. Glasserman. "Pricing american-style securities using simulation*,"* Journal of Economic Dynamics & Control, vol. 21, pp. 1323 – 1352, 1997.

[11] Rogers L. C. G. . "Monte Carlo Valuation of American Options. Mathematical Finance," *Mathematical Finance*, vol. 12, pp. 271 – 286, 1997.

[12] Longstaff F. and E.Schwartz. "Valuing American Options by Simulation: A Simple Least-Squares Approach," *The Review of Financial Studies*, vol.14, No. 1, pp. 113 – 147, 2001.

[13] Chaudhary S. "American options and the LSM algorithm: Quasirandom sequences and Brownian bridges.," preprint, 2003.

[14] NVIDIA Corporation. *CUDA programming guide*., NVIDIA Corp.OpenMP Architecture Review Board, Santa Clara, CA 2005.

[15] Sutter H. and Larus J. "Software and the concurrency revolution," *ACM Queue*, vol. 3, No. 7, pp. 54-62, 2005.

[16] David B. Kirk ,Wen-mei W. Hwu. *Programming Massively Parallel Processors: A Hands-on Approach* , Morgan Kaufmann, ISBN-10: 0123814723, pp. (10,74), 2010.

[17] Victor W Lee, Changkyu Kim, Jatin Chhugani, Michael Deisher, Daehyun Kim, Anthony D. Nguye, Nadathur Satish, Mikhail Smelyanskiy, Srinivas Chennupaty, Per Hammarlund, Ronak Singhaland Pradeep Dubey. "Debunking the 100X GPU vs. CPU Myth: An Evaluation of Throughput Computing on CPU and GPU," *37th Annual International Symposium on Computer Architecture (ISCA 2010)*, pp. 451-460, 2010.

[18] Jike Chong, Ekaterina Gonina, and Kurt Keutzer. "Monte Carlo methods: a computational pattern for our pattern language," *Proceeding ParaPLoP '10*, ACM, ISBN: 978-1-4503-0127-5, 2010.

[19] John N. Tsitsiklis and Benjamin Van Roy, "Regression Methods for Pricing Complex American-Style Options," *IEEE Transactions on Neural Networks*, vol. 12, no. 4, July.

[20] NVIDIA Corporation, CUDA C/C++ SDK CODE Samples, http://developer.nvidia.com/cuda-cc-sdk-code-samples